\documentclass{epl}

\title{Anisotropic $s$-wave superconductivity:
comparison with experiments on MgB$_2$ single crystals }

\shorttitle{Anisotropic $s$-wave superconductivity}
\author{\small A.~I.~Posazhennikova\inst{1,2}, T.~Dahm\inst{3,4},
K.~Maki\inst{4,5} }
\institute{
  \inst{1} Laboratorium voor Vaste-Stoffisica en Magnetisme, Katholieke Universiteit Leuven,
 B-3001 Leuven, Belgium\\
  \inst{2} Max-Planck-Institute for
the Chemical Physics of Solids, N\"othnitzer Str.~40, D-01187
Dresden, Germany \\
  \inst{3} Universit\"at T\"ubingen, Institut f\"ur Theoretische Physik,
Auf der Morgenstelle 14, 72076 T\"ubingen, Germany\\
  \inst{4} Max-Planck-Institute for Physics of Complex Systems,
N\"othnitzer Str.~38, D-01187 Dresden, Germany \\
  \inst{5} Department of Physics and Astronomy, University of Southern
California, Los Angeles, CA 90089-0484, USA
}

\begin{document}

\maketitle

\begin{abstract}
 The recently discovered superconductivity in MgB$_2$ has
captured world attention due to its simple crystal structure and
relatively high superconducting  transition temperature $T_c=39K$.
It appears to be generally accepted that it is phonon-mediated
$s$-wave BCS like superconductivity. Surprisingly, the strongly
temperature dependent anisotropy of the upper critical field,
observed experimentally in magnesium diboride single crystals, is
still lacking a consistent theoretical explanation. We propose a
simple single-gap anisotropic $s$-wave order parameter in order to
compare its implications with the predictions of a multi-gap
isotropic $s$-wave model. The quasiparticle density of states,
thermodynamic properties, NMR spin-lattice relaxation rate,
optical conductivity, and $H_{c2}$ anisotropy have been analyzed
within this anisotropic $s$-wave model. We show that the present
model can capture many aspects of the unusual superconducting
properties of the MgB$_2$ compound, though more experimental data
appear to be necessary from single crystal MgB$_2$.
\end{abstract}

\section{Introduction}
The moderately high temperature superconductivity with $T_c\simeq
40K$, discovered about one year ago in MgB$_2$
\cite{Akimitsu,Buzea} has stimulated an intense research activity
all over the world. Superconductivity in this binary compound
appears to be due to an electron-phonon interaction and compatible
with BCS $s$-wave superconductivity. Thermodynamics and contact
tunneling data as well as some theoretical studies indicate that
superconductivity in MgB$_2$ is one of the rare examples of two-band
superconductivity with two energy gaps, attached to different
sheets of the Fermi surface
\cite{Szabo,Wang,Bouquet,Shulga,Bascones,An,Liu,Golubov}. On the
other hand, the two gap model appears not to be able to cope with
a strongly anisotropic upper critical field in $c$-axis oriented
MgB$_2$ films and more recently in single crystals of MgB$_2$
\cite{deLima,Xu,Angst,Ott,Eltsev,Welp}. Indeed, the anisotropic
$s$-wave model can describe in principle an anisotropy of the
upper critical field and an angular dependence of $H_{c2}$
including deviations from the Ginzburg-Landau prediction with
anisotropic mass term \cite{Chen,Haas}. Indeed, some of the STM
data appear to be more consistent with an anisotropic $s$-wave
model \cite{Yeh}. However, more telling is the temperature
dependence of the ratio of $H_{c2}^{a}(t)$ ( $ t \equiv T/T_c $)
and $H_{c2}^c(t)$ where the superfix $a$ and $c$ denotes a field
oriented parallel to the $a$-axis or the $c$-axis, respectively.
The ratio $\gamma(t)(\equiv H_{c2}^{a}/H_{c2}^{c})$ increases as
temperature decreases as has been observed in Refs.
\cite{Xu,Angst,Ott}. This clearly indicates that $\Delta({\bf k})$
has to have oblate rather than the prolate form suggested earlier
\cite{Chen,Haas}. Otherwise $\gamma(t)$ would decrease with
decreasing temperature.

The object of this paper is to propose an oblate $\Delta({\bf k})$
gap anisotropy and to see whether this order parameter can describe
the two gap
features observed experimentally. Indeed choosing one adjustable
parameter, which determines the ratio of
$\Delta_{\rm min}/\Delta_{\rm max}$, we can describe reasonably well the
density of states measured by STM \cite{Yeh} and the specific heat data
of Wang et al \cite{Wang}. Also some peculiarities of the NMR relaxation
rate and optical conductivity can be described. Then we shall go on to
study the upper
critical field for ${\bf H}\parallel {\bf c}$ and ${\bf
H}\parallel {\bf a}$.

As we shall see later, we can describe $H_{c2}^c(t)$ reasonably
well by choosing our parameter $a > 5$
(corresponding to a ratio of
$\Delta_{\rm min}/\Delta_{\rm max} < 0.4$, see below).
On the other hand,
the strong increase of $H_{c2}^{a}(t)$ appears to be somewhat
difficult to fit consistently with other experimental data within our model.
Of course in the real single
crystals the electronic mean free path is rather short $l\simeq
60\sim 70$\AA \cite{Xu,Angst,Ott,Jung,Eltsev2}. Therefore in a more
realistic
analysis it is necessary to consider impurity scattering as well.

\section{Thermodynamics}

\begin{figure}
\twofigures[angle=270,width=7cm]{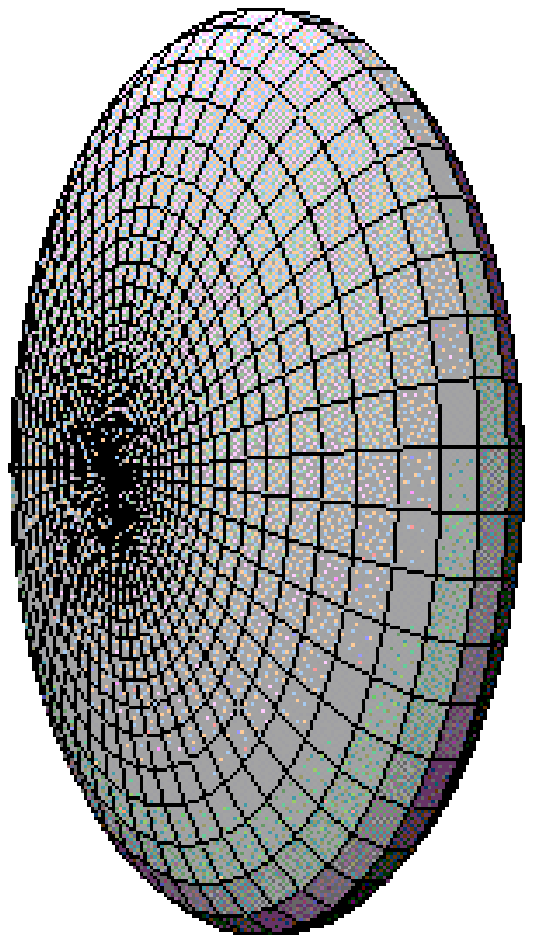}
{pgfig2.ps}
\vspace{0.3cm} \caption{Anisotropic $s$-wave order parameter Eq. (\ref{op})
for $a=10$}
\label{fig1}
\caption{Density of states for $a$= 5, 10, and 20.}
\label{fig2}
\end{figure}

We consider an anisotropic BCS model for superconductivity in MgB$_2$ with
an order parameter given by
\begin{equation}\label{op}
\Delta({\bf k})=\Delta \frac{1}{\sqrt{1+a z^2}},
\end{equation}
where the parameter $a$ determines the anisotropy,
$z=\cos(\theta)$ and $\theta$ is the polar angle with respect to
the $c$-axis. In Fig.~\ref{fig1} the anisotropic $s$-wave order
parameter is plotted in momentum space for $a=10$ displaying a
pancake-like shape. Its maximum value $\Delta_{\rm max}=\Delta$
lies in the $ab$-plane, while in $c$-axis direction it assumes
its minimum value of $\Delta_{\rm min}=\Delta/\sqrt{1+a}$. In the
following we will explore the consequences of anisotropic $s$-wave
superconductivity in MgB$_2$ within the framework of the
weak-coupling BCS theory.

\begin{table}
\caption{Gap ratios and specific heat jumps for $a$= 5, 10, 20, and 40.}
\label{t.1}
\begin{center}
\begin{tabular}{cccc}
$a$  & $\frac{\Delta_{\rm max}}{T_c}$ & $\frac{\Delta_{\rm min}}{T_c}$
& $\frac{\Delta C}{\gamma T_c}$ \\[1ex] \hline
5  & 2.29 & 0.93 & 1.11 \\
10 & 2.46 & 0.74 & 0.93 \\
20 & 2.61 & 0.57 & 0.74 \\
40 & 2.78 & 0.43 & 0.57
\end{tabular}
\end{center}
\end{table}

We have solved numerically the weak-coupling gap equation for
the state Eq. (\ref{op})
\begin{equation}\label{gapeq}
\int_0^\infty d\epsilon \left\{ \left\langle f^2(z) \right\rangle^{-1}
\left\langle \frac{f^2(z) }{\sqrt{\epsilon^2 + \Delta^2 f^2(z)}}
\tanh \left( \frac{\sqrt{\epsilon^2 + \Delta^2 f^2(z)}}{2 T} \right)
\right\rangle
- \frac{1}{\epsilon}
\tanh \left( \frac{\epsilon}{2 T_c} \right) \right\} = 0
\end{equation}
Here, $f(z)=1/\sqrt{1+a z^2}$ and $\langle \cdots \rangle=\int_0^1 dz \cdots$
denotes an angular average over the variable $z$. In Table~\ref{t.1}
we tabulated gap ratios found from the solution of
Eq. (\ref{gapeq}) for $a$= 5, 10, 20, and 40. With increasing value
of $a$ an increasing
gap ratio of the maximum gap $\Delta_{\rm max}/T_c$ is found,
while the minimum gap ratio appears to decrease.

In Fig.~\ref{fig2} we show the corresponding density of states
for these values of $a$ calculated from
\begin{equation}\label{dos}
N(E)/N_0 = {\rm Re} \; \Big\langle \frac{E}{\sqrt{E^2 - \Delta^2 f^2(z)}}
\Big\rangle
\end{equation}

The data in \cite{Yeh} resembles our density of states if we
choose $a\sim 3$. Further the peak in the density of states in
\cite{Yeh} gives a maximum gap $\sim 8$~meV consistent with our
model. Also the authors \cite{Yeh} deduced a momentum dependence
of $\Delta({\bf k})$, which is topologically the same as the
earlier model \cite{Chen,Haas}. On the other hand, the data in
\cite{Giubileo} exhibits a two gap like structure very different
from Fig.~\ref{fig2}. Further, the maximum gap $7$~meV is somewhat
smaller then the one expected from the present model. However, the
double gap like feature is inherent to the anisotropic gap, since
the two extrema associated with the single gap are always visible
\cite{Chen,Haas}.

\begin{figure}
\twofigures[angle=270,width=7cm]{pgfig3.ps}{pgfig4.ps}
\vspace{0.3cm} \caption{Reduced specific heat $C/\gamma T$ for
$a$= 10, 20, and 40. The symbols are data from Ref.~\cite{Wang}.}
\label{fig3}
\caption{Normalized NMR spin-lattice relaxation rate
$T_1^{-1}/T_{1N}^{-1}$ for $a$= 5, 10, and 20.}
\label{fig4}
\end{figure}

In Fig.~\ref{fig3} the specific heat is shown for values of
$a$= 10, 20, and 40
along with the data of Wang et al. \cite{Wang,Bouquet}. The specific heat
jumps found from our model are also tabulated in  Table~\ref{t.1}.
They decrease with increasing $a$.
The closest agreement with the experimental data is obtained for $a=20$.
However, fitting of the structures seen in the experimental data would
require some fine tuning of the angular dependence in Eq. (\ref{op}).

In Fig.~\ref{fig4} we show the normalized NMR spin-lattice relaxation
rate obtained from our state. With increasing anisotropy $a$ the
Hebel-Slichter peak is reduced. Roughly, the size of the
peak found here is consistent with recent NMR data on
polycrystalline MgB$_2$ by Kotegawa et al. \cite{Kotegawa}.

It has recently been noted that the gap seen in optical conductivity
appears to be very small \cite{Kaindl,Pimenov}. We calculated the
anisotropic analog of the Mattis-Bardeen conductivity \cite{Mattis}
averaged over all directions, which should correspond to the
response of a polycrystalline thin film sample as has been used in Refs.
\cite{Kaindl,Pimenov}. For our anisotropic state the conductivity
$\sigma(\omega)$ is obtained using the following two
functions:
\[
I_1(x)  =  \int_0^1 dz \int_{f(z)}^\infty dy
\frac{ f^2(z) + y (y+x)}
{\sqrt{\left( y^2 - f^2(z) \right) \left[ (y+x)^2 - f^2(z) \right] }}
\left[ \tanh \left( \frac{(x+y) \Delta}{2 T} \right) -
\tanh \left( \frac{ y \Delta}{2 T} \right) \right]
\]
\[
I_2(x,z_0)  =  \int_{z_0}^1 dz
\int_{f(z)}^\infty dy \frac{ f^2(z) - y (x-y) }
{\sqrt{\left( y^2 - f^2(z) \right) \left[ (y-x)^2 - f^2(z) \right] }}
\tanh \left( \frac{ y \Delta}{2 T} \right)
\]
Using these the optical conductivity reads
\begin{equation}
\frac{\sigma (\omega)}{\sigma_N} = \left\{
\begin{array}{c@{\quad {\rm for} \quad}c}
\frac{\Delta}{\omega} I_1 \left( \frac{\omega}{\Delta} \right) &
\omega < 2 \Delta_{\rm min} \\
\frac{\Delta}{\omega} \left[ I_1 \left( \frac{\omega}{\Delta} \right) -
I_2 \left( \frac{\omega}{\Delta} ,
\sqrt{\frac{4 \Delta^2-\omega^2}{a \omega^2}} \right) \right] &
2 \Delta_{\rm min} < \omega < 2 \Delta_{\rm max} \\
\frac{\Delta}{\omega} \left[ I_1 \left( \frac{\omega}{\Delta} \right) -
I_2 \left( \frac{\omega}{\Delta} , 0 \right) \right] &
\omega > 2 \Delta_{\rm max}
\end{array} \right.
\end{equation}
The result for $a=20$ and different reduced temperatures is shown
in Fig.~\ref{fig5}. Indeed it appears that the optical
conductivity is mainly dominated by the minimum gap within our
model, because the absorption threshold starts at $2 \Delta_{\rm
min}$. Thus, an anisotropic gap scenario can also account for the
small gap seen in the optical conductivity. Most useful would be
optical conductivity studies on single crystals which could
resolve the $c$-axis and $ab$-plane response in order to see more
clearly, what the anisotropy looks like.

\begin{figure}
\twofigures[angle=270,width=7cm]{pgfig5.ps}{pgfig6c.ps}
\vspace{0.3cm} \caption{Mattis-Bardeen conductivity as a function
of normalized frequency $\omega/\Delta$ for $a=20$ and reduced
temperatures $t=T/T_c$ of 0.9, 0.8, 0.6, 0.4, and 0.2 (from top to
bottom), respectively.} \label{fig5} \caption{Normalized upper
critical field $h_{c2}$ as a function of reduced temperature $t$
for $a=10$ and $a=100$. The upper two curves are for field
direction within the $ab$-plane, while the lower two curves are
for field in $c$-axis direction. For comparison also the isotropic
result $(a=0)$ is shown. The data points are taken from Ref.
\cite{Angst}, where we have normalized them to the slope at $T_c$,
taking $T_c=37.3$K, $\frac{d H_{c2,ab}}{d
T}|_{T_c}\simeq-0.435$T/K, and $\frac{d H_{c2,c}}{d
T}|_{T_c}\simeq-0.124$T/K.  } \label{fig6}
\end{figure}

\section{Upper critical field}

Recent experiments on the upper critical field in MgB$_2$ single
crystals have reported a strong temperature dependent anisotropy
of $H_{c2}$ and an unusual upward curvature of the critical field
parallel to the $ab$ plane. An anisotropy of the superconducting
pairing can in principle give such an effect. We investigate the
temperature dependence of both $H_{c2}$, parallel and
perpendicular to the crystal $c$-axis of MgB$_2$.

The general equation for the upper critical field is derived from
the gap equation. For unconventional superconductors it can be
treated variationally, some technical details can be found
elsewhere \cite{Sun}.

\noindent
 (a) ${\bf H}\parallel {\bf c}$

The equation for the upper critical field parallel to the $c$-axis
is given by
\begin{equation}\label{hc2parc}
-\ln{t}=\int_0^{\infty}
\frac{du}{\sinh{u}}\Bigg(1-\langle f^2 \rangle^{-1}\Big\langle
\exp(-\rho u^2 (1-z^2))f^2\Big\rangle \Bigg),
\end{equation}
where $\rho=v_a^2eH_{c2}(t)/2(2\pi T)^2$, and $v_a$ is the Fermi
velocity within the $ab$-plane.


In Fig.~\ref{fig6} the numerical solution of this equation,
normalized by its derivative at $T_c$, $h_{c2}(T)\equiv
H_{c2}(T)/(-T_c\partial H_{c2}(T)/\partial T)|_{T_c}$, is plotted
along with the $h_{c2}(T)$-curves for the $ab$-direction, the
corresponding curve for the isotropic case, and normalized
experimental data by Angst et al. \cite{Angst}. In the limit
$T\rightarrow 0$ for $a=10$ we have $h_{c2}^c(0)\simeq 0.63349$ which is
somewhat smaller than the corresponding value in the isotropic
$s$-wave superconductor $h_{c2}(0)\simeq 0.72726$. In the vicinity
of $T_c$ the upper critical field exhibits naturally a rather linear
temperature dependence within the weak-coupling BCS theory. This
behavior is observed for both anisotropic $s$-wave and
conventional superconductors. Note, that our results for the upper
critical field parallel to the $c$-axis for $a>5$
are in good correspondence with the experimental data \cite{Angst}.
The best fit to these data is obtained for $a \approx 10$. However,
we cannot reliably extract a value of $a$ from these data, because
$h_{c2}^c(0)$ for $a \rightarrow \infty$ is saturating at a
value of 0.59054 and the data are already close to this limit.
Thus, also even higher values of $a$ might still be consistent
with the $c$-axis data.

\noindent
(b) ${\bf H}\parallel {\bf a}$

The derivation of the equation for the temperature dependent
upper critical field in the plane is more involved. The problem is
that a mixing of higher Landau levels takes place \cite{Sun}. We
choose a variational wavefunction suggested in
Ref.~\cite{Schopohl}, Eq. (36), which corresponds to a distorted
Abrikosov ground state. Due to the anisotropy of our state such a
distortion is expected for the field direction perpendicular to
the $c$-axis. The distortion can be varied using a parameter
$\alpha$, which has to be determined by a variational principle.
Using this method we arrive at the following equation for the
temperature dependence of the upper critical field
$\rho_a(t)=v_av_ceH_{c2}(t)/2(2\pi T)^2$
\begin{equation}
-\ln{t}  =  \int_0^{\infty} \frac{d u}{\sinh{u}}
\Bigg\{ 1-\left\langle
f^2\right\rangle^{-1} \left\langle
\exp \left\{ - \rho_a u^2 \left( \frac{1}{\alpha^2} (1-z^2) \cos^2 \phi +
\alpha^2 z^2 \right) \right\} f^2
\right\rangle \Bigg\},
\label{eqrhoa}
\end{equation}
here $\langle \cdots \rangle=\int_0^1dz\int_0^{2\pi}\frac{d
\phi}{2\pi} \cdots$ and $v_c$ is the Fermi velocity in $c$-axis
direction. The parameter $\alpha$ has to be determined from a
minimization of the right-hand side of
Eq.~(\ref{eqrhoa}), taking its derivative to be zero.

We have compared our results with a second, independent calculation
based on a Landau level expension up to the fourth Landau level
following the method of Luk'yanchuk and Mineev
\cite{Mineev}. For the parameter range $a<20$ both methods were in
close agreement with each other. However, for $a \gg
20$ the Landau level expansion started to give significantly smaller
values for $H_{c2}^{a}$, indicating that
inclusion of even higher Landau levels becomes necessary, and
showing that the distorted Abrikosov wavefunction is a better
variational solution in this case.

The numerical solution of Eq.~(\ref{eqrhoa}) obtained from a
numerical optimization of $\alpha$ is shown in Fig.~\ref{fig6}. In
the zero-temperature limit we obtain $h_{c2}^{a} = 0.87309$ for
$a=10$ and $h_{c2}^{a} = 1.18025$ for $a=100$. The curve for
$a=100$ possesses a slight upward curvature in agreement with the
experimental data. From our comparison with the experimental data
\cite{Angst} we extracted the values $-\partial
H_{c2}^{a}(T)/\partial T \simeq 0.435$~T/K and $-\partial H_{c2}^c
(T)/\partial T\simeq 0.124$~T/K. Choosing $a=10$, from these we
deduce average Fermi velocities $v_a\simeq 2.7 \cdot10^7$~cm/sec,
$v_c\simeq 1.3 \cdot10^7$~cm/sec and the ratio $v_c/v_a\simeq
0.48$. The last ratio is very consistent with what was obtained
earlier \cite{Chen} using the data in \cite{deLima}. Also the
absolute values of these Fermi velocities $v_a$ and $v_c$ are
roughly consistent with bandstructure calculation results provided
to us by A.~Yaresko.

This rather large value of $a=100$ appears to give a reasonable
fit to the $ab$-plane data of $H_{c2}$. But unfortunately, this
corresponds to a ratio of $\Delta_{\rm min}/\Delta_{\rm max}
\simeq 0.1$, which appears to be inconsistent with the values we
discussed above. Here we can just speculate about a possible
resolution of this problem: in these single crystals the mean free
path is relatively short, being about the order of the coherence
length \cite{Jung,Eltsev2}, which puts these systems in an
intermediate region between the clean and the dirty limit.
Therefore a study of the influence of impurity scattering would be
necessary. It is possible that the upper critical field anisotropy
is much less sensitive to impurity scattering than the energy gap
and the density of states, for example. Such an effect could
possibly resolve this problem and is left for a future work.

\section{Conclusions}
We have described a model of anisotropic s-wave superconductivity
and compared our theoretical predictions with the experimental
data of MgB$_2$. We show that our model can capture many aspects
of the two gap model and is consistent with the experimental data
of single crystals MgB$_2$. Also we stress that so far our model
is the only one, that can describe the strongly temperature
dependent anisotropy in the upper critical field. Unfortunately,
strongly different values of our anisotropy parameter $a$ had to
be used for different experimental properties. On the other hand,
we have ignored the effect of impurity scattering in the present
analysis. In particular the mean free path $l\sim
60-70$\AA~reported for single crystals of MgB$_2$ implies that we
are in the intermediate regime (neither in the clean limit nor in
the dirty limit). Therefore in a more realistic analysis, it will
be very important to incorporate the effect of impurity
scattering. Clearly, we also need more precise measurements on
single crystal MgB$_2$.

\acknowledgments We would like to thank M.~Angst, F.~Bouquet, and
A.~Junod for providing us with the digital form of their
experimental data. Thanks are also due to Y.~Takano, J.~Karpinski,
P.~Thalmeier, A.~Yaresko, and N.~Schopohl for useful discussions
related to this subject. We are indebted to Todor M.~Mishonov
for pointing out to us an error in Table~\ref{t.1} in the
first version of this manuscript.

\end{document}